\documentclass[sigconf]{acmart}

\usepackage{amsthm}
\usepackage{bm}
 
\setcopyright{acmlicensed}
\copyrightyear{2026}
\acmYear{2026}
\acmDOI{XXXXXXX.XXXXXXX}

\acmConference[PASC '26]{Platform for Advanced Scientific Computing Conference}{June 29--July 1, 2026}{Bern, Switzerland}
\acmISBN{978-1-4503-XXXX-X/2026/06}

\newcommand{\mc}[1]{\mathcal{#1}}
\newcommand{\mb}[1]{\bm{#1}}

\newcommand{\LRp}[1]{\left( #1 \right)}
\newcommand{\LRs}[1]{\left[ #1 \right]}

\newcommand{\eqnlab}[1]{\label{eq:#1}}
\newcommand{\eqnref}[1]{\eqref{eq:#1}}

\newcommand{\xb}{\mb{x}}
\newcommand{\vb}{\mb{v}}
\newcommand{\fb}{\mb{f}}
\newcommand{\kb}{\mb{k}}
\newcommand{\ib}{\mb{i}}

\newcommand{\fbk}{\mb{f}_{\mb{k}}}

\newcommand{\Cb}{\mb{C}}
\newcommand{\Db}{\mb{D}}
\begin{document}

\title{A Performance-Portable, Massively Parallel Distributed Nonuniform FFT}

\author{Paul Fischill}
\email{pfischill@ethz.ch}
\orcid{0009-0005-9850-9415}
\affiliation{
  \institution{ETH Z\"urich}
  \city{Z\"urich}
  \country{Switzerland}}

\author{Andreas Adelmann}
\email{andreas.adelmann@psi.ch}
\orcid{0000-0002-7230-7007}
\affiliation{
 \institution{PSI Center for Scientific Computing, Theory and Data}
 \city{Villigen PSI}
 \country{Switzerland}}

\author{Sriramkrishnan Muralikrishnan}
\email{s.muralikrishnan@fz-juelich.de}
\orcid{0000-0002-5494-1880}
\affiliation{
  \institution{J\"ulich Supercomputing Centre, Forschungszentrum J\"ulich GmbH}
  \city{J\"ulich}
  \country{Germany}}

\renewcommand{\shortauthors}{Fischill et al.}

\begin{abstract}
The nonuniform fast Fourier transform (NUFFT) enables spectral methods for problems with irregularly spaced samples, with applications in medical imaging, molecular dynamics, and kinetic plasma simulations. Existing implementations are limited to shared-memory execution, restricting problem sizes to what fits on a single node. We present the first distributed, performance-portable NUFFT for heterogeneous supercomputers. Our Kokkos-based implementation runs without modification on NVIDIA and AMD GPUs. We develop multiple spreading and interpolation kernels optimized for different accuracy requirements and architectures. Our spreading kernels match or exceed the single-GPU throughput of the state-of-the-art CUDA-based NUFFT library cuFINUFFT at production particle densities, while our Kokkos-based implementation additionally supports AMD GPUs. Strong scaling experiments on Alps (NVIDIA GH200), JUWELS Booster (NVIDIA A100), and LUMI (AMD MI250X) demonstrate scaling up to 1024 GPUs. At scale, the distributed FFT is a significant part of the total runtime, making higher NUFFT accuracy less expensive. We apply the method to massively parallel Particle-in-Fourier simulations of Landau damping with up to $1024^3$ Fourier modes and 8.6 billion particles on Alps, JUWELS, and LUMI, demonstrating that distributed NUFFTs enable kinetic plasma simulations at resolutions previously inaccessible to spectral particle methods.
\end{abstract}



\maketitle

\section{Introduction}

The nonuniform fast Fourier transform (NUFFT)~\cite{dutt_fast_1993,dutt_fast_1995} computes Fourier transforms between regularly spaced grids and irregularly distributed samples in $\mathcal{O}(N \log N + M)$ time, where $N$ is the grid size and $M$ the number of nonuniform points. Applications span magnetic resonance imaging~\cite{fessler_nonuniform_2003}, molecular dynamics~\cite{arnold_comparison_2013}, kinetic plasma simulations~\cite{mitchell_efficient_2019,shen_particle--fourier_2024,muralikrishnan_error_2025}, and fluid-structure interaction~\cite{chen_fourier_2024}. Further examples appear in~\cite{barnett_parallel_2019,shih_cufinufft_2021,potts_fast_2001,greengard_accelerating_2004}.

Several open-source NUFFT libraries cover parts of this design space.
FINUFFT~\cite{barnett_parallel_2019} provides a high-performance C
implementation for CPUs, and cuFINUFFT~\cite{shih_cufinufft_2021}
adds support for NVIDIA GPUs. NFFT3~\cite{keiner_using_2009} is a
CPU-only C library, while CUNFFT~\cite{kunis_nonequispaced_2012}
provides a CUDA implementation. gpuNUFFT~\cite{knoll_gpunufft_2014}
targets cryo-EM workloads on NVIDIA hardware. In Julia,
NonuniformFFTs\allowbreak.jl~\cite{Polanco_NonuniformFFTs_jl_2024} supports CPUs
and multiple GPU backends. Python packages include PyNUFFT~\cite{lin_python_2018} 
and \texttt{TorchKbNufft}~\cite{muckley:20:tah}, the
latter using PyTorch for multi-vendor GPU support.

While Julia and PyTorch libraries offer vendor portability, most production particle codes in scientific computing are written in C/C++/Fortran and require interfaces to low-level libraries. The widely adopted NFFT3 remains CPU-only, 
while FINUFFT supports CPUs and NVIDIA GPUs.

More critically, all of the aforementioned libraries support only shared-memory parallelism, restricting execution to a single node or GPU. This limits problem size to the memory of one device, a significant constraint for molecular dynamics~\cite{arnold_comparison_2013}, fluid-structure interaction~\cite{chen_fourier_2024}, and kinetic plasma simulations~\cite{mitchell_efficient_2019,shen_particle--fourier_2024,muralikrishnan_error_2025}, where high-fidelity simulations require billions of Fourier modes and particles. The only distributed NUFFT implementation we are aware of is PNFFT~\cite{pippig_parallel_2013,pippig_massively_2016}, which has enabled billion-particle molecular dynamics simulations~\cite{arnold_comparison_2013}. However, PNFFT is CPU-only and has not been actively maintained for over a decade.

These limitations motivate a distributed NUFFT library that is
performance-portable across heterogeneous supercomputers.

\paragraph{Contribution}

We present what is, to the best of our knowledge, the first distributed, performance-portable NUFFT implementation. Our library scales efficiently on leading European supercomputers including Alps (NVIDIA GH200), JUWELS Booster (NVIDIA A100), and LUMI (AMD MI250X). The implementation builds on the Independent Parallel Particle Layer (IPPL)~\cite{muralikrishnan2024scaling,matthias_frey_ippl-frameworkippl_2024}, an open-source C++ framework for large-scale particle-mesh methods. IPPL leverages Kokkos~\cite{carter_edwards_kokkos_2014} for on-node performance portability, MPI for distributed parallelism, and HeFFTe~\cite{krzhizhanovskaya_heffte_2020} for distributed FFTs.

\section{NUFFT}

The nonuniform discrete Fourier transform (NUDFT) is the direct Fourier transform between irregularly spaced sample locations and a uniform set of Fourier modes~\cite{dutt_fast_1993,dutt_fast_1995,potts_fast_2001,greengard_accelerating_2004}. In this work we consider the two standard variants used in NUFFT algorithms. Type~1 maps data given at nonuniform sample locations to uniformly spaced Fourier coefficients, while Type~2, which is the adjoint of Type~1, maps uniformly spaced Fourier coefficients back to values at nonuniform sample locations~\cite{dutt_fast_1995,dutt_fast_1993,barnett_parallel_2019}. 

Let $\fb(\xb_j)$ denote the value associated with the $j$th nonuniform sample located at position $\xb_j \in [0,L)^3$, and let $\fbk(\kb)$ denote the Fourier coefficient associated with mode $\kb$. We write $K_N$ for the set of retained Fourier wavevectors on an $N \times N \times N$ uniform Fourier grid. With this notation, the Type~1 and Type~2 transforms are
\begin{alignat}{2}
\eqnlab{p2f}
\fbk(\kb)
  &= \sum_{j=1}^{N_p} \fb(\xb_j)e^{-\mathrm{i}\kb\cdot\xb_j},
  &\qquad \kb &\in K_N, \\
\eqnlab{f2p}
\fb(\xb_j)
  &= \sum_{\kb\in K_N} \fbk(\kb)e^{\mathrm{i}\kb\cdot\xb_j},
  &\qquad j &= 1,\ldots,N_p.
\end{alignat}
For simplicity of exposition, we restrict attention to the three-dimensional case
and follow the standard NUFFT formulation~\cite{dutt_fast_1993,dutt_fast_1995}.
Let $N$ denote the number of Fourier modes retained in each coordinate direction,
so that the total number of retained modes is $N_m = N^3$. Let $L$ be the domain
length in each direction. Throughout, we assume that $N$ is even and define
\[
K_N =
\left\{
\frac{2\pi}{L}\,\bm{n}
:
\bm{n}\in\{-N/2,\ldots,N/2-1\}^3
\right\}.
\]
Thus, $K_N$ is the set of all discrete wavevectors represented on the uniform Fourier grid, while $\xb_j$ is the position of sample $j$ and $N_p$ is the total number of samples. A direct evaluation of either transform requires $\mc{O}(N_pN_m)$ work, since each of the $N_p$ nonuniform samples interacts with each of the $N_m$ Fourier modes~\cite{dutt_fast_1993,dutt_fast_1995,potts_fast_2001}. This cost is prohibitively expensive except for very small numbers of samples or Fourier modes.

In Equation~\eqnref{p2f}, each Fourier coefficient $\fbk(\kb)$ is obtained by summing the contributions from all nonuniform samples, each weighted by the complex exponential associated with wavevector $\kb$. Thus, Type~1 may be viewed as a ``scatter to Fourier space'' operation. In Equation~\eqnref{f2p}, the value at a sample location $\xb_j$ is reconstructed by summing all retained Fourier modes with the corresponding inverse phase factor. Type~2 therefore acts as a ``gather from Fourier space'' operation. The two transforms are adjoint to one another under the standard discrete inner product~\cite{dutt_fast_1993,dutt_fast_1995,potts_fast_2001,keiner_using_2009,barnett_parallel_2019}.

The NUFFT accelerates the direct transforms in Equations~\eqnref{p2f} and~\eqnref{f2p} by replacing the dense interaction between all particles and all Fourier modes with three structured operations: spreading or interpolation between particles and an oversampled uniform grid, a standard FFT on that grid, and a diagonal deconvolution (or pre-correction) step~\cite{potts_fast_2001,greengard_accelerating_2004,barnett_parallel_2019}. This reduces the complexity from $\mc{O}(N_pN_m)$ to $\mc{O}(|\log \varepsilon|^d N_p + N_m \log N_m)$, where $\varepsilon$ is the requested NUFFT tolerance and $d$ is the spatial dimension~\cite{potts_fast_2001,barnett_parallel_2019}. In matrix form, the Type~1 NUFFT can be written as
\begin{equation}
\eqnlab{type1nufft}
    \fbk =
    \Db_{N_m}
    \chi_{N_m,N_{\mathrm{os}}}
    \mc{F}_{N_{\mathrm{os}},N_{\mathrm{os}}}
    \Cb_{N_{\mathrm{os}},N_p}
    \fb .
\end{equation}
This factorization follows the standard operator formulation of window-based
NUFFTs~\cite{potts_fast_2001,keiner_using_2009}. The vector $\fb$ contains the
data values at the nonuniform sample locations, and $\fbk$ contains the
corresponding Fourier coefficients on the uniform mode grid. The computation
consists of the following steps.
\begin{itemize}
    \item  \textbf{Step 1:} The data $\fb$ is spread onto an oversampled grid of size $N_{\mathrm{os}} = M^3$ where $M = \sigma N$ and $\sigma > 1$. Typically, $\sigma=2$ is selected to 
    reduce aliasing~\cite{barnett_parallel_2019,barnett_aliasing_2021}. The spreading is performed using a window function $\varphi$ that is well-localized in both physical and frequency space, so that it has compact support during spreading while introducing only controlled aliasing in Fourier space~\cite{potts_fast_2001,barnett_parallel_2019,barnett_aliasing_2021,potts_uniform_2021,potts_continuous_2021}. In Equation~\eqnref{type1nufft}, the spreading operation is denoted by the real, sparse matrix $\Cb_{N_{\mathrm{os}},N_p}$ which accepts as input the sample data of size $N_p$ and gives as output the data on the grid of size $N_{\mathrm{os}}$.
   
\item \textbf{Step 2:} A uniform FFT is applied to the data on the oversampled grid. This operation is denoted by $\mc{F}_{N_{\mathrm{os}},N_{\mathrm{os}}}$ in~\eqnref{type1nufft}.

    \item \textbf{Step 3:} From the FFT output, the $N_m$ Fourier modes corresponding to $K_N$ are extracted using the characteristic function matrix $\chi_{N_m,N_{\mathrm{os}}}$ which has ones in the 
    locations of the modes to be selected and zeros in all other places.

    \item \textbf{Step 4:} Finally, the smoothing introduced by the window function is removed by deconvolution.  
    $\Db_{N_m}$ is a diagonal matrix of deconvolution factors. Its entries are
    proportional to $1/\hat{\varphi}(\kb)$, including the normalization constants
    associated with the FFT convention, for $\kb \in K_N$.

\end{itemize}

For the Type~2 NUFFT in Equation~\eqnref{f2p}, these steps are reversed,
giving the matrix expression
\begin{equation}
\eqnlab{type2nufft}
    \fb  = \Cb^{\top}_{N_p,N_{\mathrm{os}}} \mc{F}^{-1}_{N_{\mathrm{os}},N_{\mathrm{os}}}\chi^\top_{N_{\mathrm{os}},N_m} \Db_{N_m} \fbk.
\end{equation}

\subsection{Window Functions and Kernel Design}

The NUFFT algorithm requires a window function $\varphi$ that is well-localized in both spatial and frequency domains. Several window functions have been proposed and proven optimal or quasi-optimal in the sense of minimal support for a given tolerance~\cite{potts_uniform_2021,potts_continuous_2021}. These include the Kaiser-Bessel function~\cite{kaiser1966digitalfilters}, the prolate spheroidal wave function (PSWF)~\cite{osipov_prolate_2013}, the Gaussian~\cite{dutt_fast_1993}, and the exponential-of-semicircle (ES) kernel~\cite{barnett_parallel_2019}.

We adopt the ES kernel, defined as
\begin{equation}
    \varphi(x) = \begin{cases}
        e^{\beta(\sqrt{1 - x^2} - 1)}, & |x| \leq 1, \\
        0, & \text{otherwise},
    \end{cases}
\end{equation}
where $\beta > 0$ controls the trade-off between spatial localization and frequency decay. The ES kernel achieves error rates close to the optimum while being significantly simpler to evaluate~\cite{barnett_parallel_2019,barnett_aliasing_2021,potts_continuous_2021}.

We evaluate $\varphi$ directly at runtime rather than using polynomial approximations~\cite{shih_cufinufft_2021}. While polynomial fitting reduces arithmetic operations, our measurements in Section~\ref{sec:performance_eval} show that both spreading and interpolation are memory-bandwidth limited on modern GPUs. Consequently, reducing computation provides negligible performance gains.

The deconvolution ($\Db$) step requires the Fourier transform $\hat{\varphi}$ of the window function in order to form the reciprocal correction factors.
For the ES kernel, no closed-form expression exists, so we compute $\hat{\varphi}$ numerically via Gauss-Legendre quadrature to machine precision. This computation is performed once during initialization and reused across all subsequent transforms with the same grid shape. The cost is negligible compared to the spreading and FFT steps.

The kernel width $w$ and upsampling factor $\sigma$ are chosen based on the user-specified tolerance $\varepsilon$. Following~\cite{barnett_parallel_2019}, we use $\sigma = 2$ and select $w$ such that the aliasing error satisfies the tolerance. 

\subsection{Spreading Implementation}

The spreading operation ($\Cb$) maps values at nonuniform particle positions onto a regular grid. This is analogous to higher-order charge deposition in particle-in-cell methods~\cite{birdsall_plasma_1991,hockney_computer_1981}, and is also the dominant kernel in many practical NUFFT implementations~\cite{barnett_parallel_2019,shih_cufinufft_2021}. The main computational challenge is write conflict: when parallelizing over particles, multiple threads may attempt to update the same grid location simultaneously, making memory-system behavior and atomic contention central performance concerns on GPUs~\cite{shih_cufinufft_2021}.

All spreading kernels share a common structure. For each particle at
position $\xb_j$, we identify a base grid index $\ib_0$ and define
$a^{(d)} = i_0^{(d)} - \lceil (w-1)/2 \rceil$, so that the window
function has support on grid points
$[a^{(d)}, \, a^{(d)} + w - 1]$ in each dimension~$d$. Each particle
thus contributes to $w^d$ grid values. Since the kernel is separable, i.e.,
\begin{equation}
    \varphi(\xb - \ib) = \prod_{k=1}^{d} \varphi(x^{(k)} - i^{(k)}),
\end{equation}
only $dw$ kernel evaluations are required per particle rather than $w^d$.

To provide competitive performance across architectures, we implement multiple spreading algorithms with different trade-offs between simplicity and atomic contention. All implementations support real and complex-valued inputs with automatic dispatch based on data types.

\paragraph{Atomic Spread}
The simplest approach, used in early GPU NUFFT implementations such as CUNFFT~\cite{kunis_nonequispaced_2012}, parallelizes directly over particles. Each thread loads one particle, evaluates the $w^d$ kernel contributions in local memory, and performs atomic additions to the $w^d$ affected grid points in global memory. This algorithm is straightforward to implement but suffers from high atomic contention when particles cluster spatially. 

\paragraph{Tiled Spread}
To reduce global memory contention, we first sort particles into spatial tiles of tunable size, following the shared-memory binning strategy introduced in cuFINUFFT~\cite{shih_cufinufft_2021} and related to the sector-based approach of gpuNUFFT~\cite{knoll_gpunufft_2014}. A team of threads then spreads particles within a tile into a shared memory histogram of size $\prod_{i=1}^{d}(T_i + w)$ to accommodate stencil overlap. Threads perform atomic additions to the shared histogram. After all particles are processed, the accumulated values are flushed to global memory with one atomic operation per grid point.

In three dimensions, shared memory requirements grow rapidly with tile size. To maintain sufficient parallelism without exceeding shared memory capacity, we subdivide work along the $z$ dimension of the stencil: multiple teams process the same tile but operate on disjoint $z$-slices of the output. This increases concurrent teams by a factor proportional to $w$ without increasing shared memory requirements per team.

\paragraph{Grid-Parallel Spread}
This approach uses the same spatial tiling and sorting as Tiled Spread, but changes the parallelization strategy within each team. Instead of parallelizing over particles and using shared memory atomics, we loop sequentially over particles and parallelize over the $w^d$ output grid points. This reversal is analogous to the grid-driven gridding introduced for MRI reconstruction~\cite{yang_cuda-based_2013,feng_cuda_2015}, and to the output-driven approach with compact binning of~\cite{gai_more_2013}. Each thread is assigned a distinct subset of the stencil, ensuring no write conflicts during accumulation. The approach requires threads to coordinate kernel evaluations via shared memory, but avoids atomics within shared memory.

\paragraph{Implementation Details}
The kernel width $w$ is a compile-time template parameter, enabling loop unrolling and register allocation optimizations. A dispatcher mechanism selects the appropriate specialization at runtime based on the requested tolerance. Particles near domain boundaries spread into ghost regions of the local grid. Contributions are then combined via halo exchange, providing uniform handling of both periodic boundary conditions and distributed-memory decomposition.

The Tiled and Grid-Parallel kernels expose several runtime parameters—tile dimensions $(T_1, T_2, T_3)$, thread team size, and $z$-axis oversubscription factor—that interact non-trivially and whose optima vary across GPU architectures. We select these parameters per platform using Bayesian optimization with a Gaussian process surrogate model over the five-dimensional configuration space, using expected improvement as the acquisition function. The optimized configurations are stored in lookup tables indexed by kernel width and loaded at runtime. The optimal algorithm depends on problem characteristics and hardware. We evaluate these trade-offs quantitatively in Section~\ref{subsec:single_gpu_results}.

\subsection{Interpolation Implementation}

The interpolation operation ($\Cb^\top$) evaluates the grid at nonuniform particle positions, analogous to higher-order particle gathering in particle-in-cell methods~\cite{birdsall_plasma_1991,hockney_computer_1981}. The structure mirrors spreading: for each particle, we identify the $w^d$ grid points with kernel support and accumulate their weighted contributions. Kernel separability reduces the cost to $dw$ kernel evaluations per particle.

Unlike spreading, interpolation involves no write conflicts. Each particle reads from the grid and writes only to its own output location. The primary challenge is memory access efficiency: achieving high cache utilization requires that concurrently executing threads access spatially coherent grid regions. We implement two algorithms:

\paragraph{Direct Interpolation}
The simplest approach parallelizes directly over particles. Each thread loads the $w^d$ grid values in its stencil, evaluates the kernel weights, and accumulates the result. This approach is simple but suffers from poor cache utilization.

\paragraph{Sorted Interpolation}
To improve cache reuse, we sort particles by spatial locality before interpolation. We provide two sorting strategies: Morton code ordering~\cite{morton1966computer}, which maps multidimensional positions to a space-filling curve, and bin sorting, which groups particles into spatial tiles~\cite{shih_cufinufft_2021}. Both approaches increase the likelihood that concurrently executing threads access nearby grid points, improving cache hit rates.

\subsection{Distributed Implementation}

We distribute the NUFFT across multiple ranks using domain decomposition, building on the distributed infrastructure provided by IPPL~\cite{matthias_frey_ippl-frameworkippl_2024}. Both grid values and particles are partitioned according to the same spatial decomposition, ensuring that each rank owns the grid region containing its particles.

The deconvolution step ($\Db$) applies pointwise scaling to grid values and requires no communication. The spreading and interpolation operations, however, access grid points within a $w^d$ stencil around each particle. Particles near subdomain boundaries therefore require access to grid points owned by neighboring ranks. We extend each local domain with a halo region of width $\lceil w/2 \rceil$ in each direction to accommodate this stencil overlap. For interpolation, halo regions must be filled with boundary values from neighboring ranks before the interpolation begins. For spreading, particles near boundaries write to halo regions. These contributions are then accumulated into owning ranks.

The distributed FFT is performed using HeFFTe~\cite{krzhizhanovskaya_heffte_2020}, which implements scalable multi-dimensional transforms across distributed memory. We integrate a pruned FFT that merges zero-padding and truncation with the transform, as described in Section~\ref{sub:pruned_fft}. This optimization is particularly beneficial in the distributed setting, where it enables proper distribution of the output values without transmitting zeros.

\subsection{Pruned FFT} \label{sub:pruned_fft}

The NUFFT requires an oversampled grid of size $M=\sigma N$, but only $N$ Fourier modes are needed. A standard approach computes the full FFT and discards unwanted modes. Pruned FFTs that skip unnecessary computation are well known in the signal processing literature~\cite{markel_fft_1971,sorensen_efficient_1993}, and partial-transform ideas appear in various FFT contexts. However, to the best of our knowledge, this technique has not previously been integrated into a distributed NUFFT pipeline, where it offers a particular advantage: the sub-transforms operate on smaller data volumes, reducing the all-to-all communication that dominates distributed FFT cost. We describe our realization for $\sigma = 2$, which extracts only the $N/2$ positive and $N/2$ negative modes (Type~1), or zero-pads these modes to size $M$ (Type~2).

We achieve pruning by performing one Cooley-Tukey decomposition step on top of the distributed FFT. The DFT of a sequence $\{x_n\}_{n=0}^{M-1}$ decomposes as
\begin{equation}
    X_k = \underbrace{\sum_{m=0}^{M/2-1} x_{2m} \, e^{-2\pi \mathrm{i} km/(M/2)}}_{E_k} + e^{-2\pi \mathrm{i} k/M} \underbrace{\sum_{m=0}^{M/2-1} x_{2m+1} \, e^{-2\pi \mathrm{i} km/(M/2)}}_{O_k},
\end{equation}
where $E_k$ and $O_k$ are size-$M/2$ DFTs over even- and odd-indexed elements. We only need modes $k \in \{0,\ldots,N/2-1\} \cup \{M-N/2,\ldots,M-1\}$, equivalently $k \in \{0,\ldots,M/4-1\} \cup \{3M/4,\ldots,M-1\}$ for $\sigma=2$, hence we compute two half-size FFTs and combine with twiddle factors $e^{-2\pi \mathrm{i} k/M}$. This generalizes to $d$ dimensions with $2^d$ independent FFTs of size $(M/2)^d$. Zero-padding reverses the process: pre-multiply inputs by conjugate twiddle factors, compute inverse FFTs, and interleave outputs.

The $2^d$ sub-FFTs execute concurrently using separate CUDA/HIP streams and independent MPI communicators on different OpenMP threads, enabling overlap of computation and communication. We expose the concurrency level as a tunable parameter to balance parallelism against resource contention.
\section{Performance Evaluation}
\label{sec:performance_eval}

We evaluate our implementation across multiple aspects: single-GPU kernel performance, comparison with a state-of-the-art library, and distributed scaling. First, we characterize the spreading and interpolation kernels, identifying optimal configurations across architectures. We then compare against the state-of-the-art NUFFT library for NVIDIA GPUs, cuFINUFFT~\cite{shih_cufinufft_2021}, to validate single-node competitiveness. Finally, we demonstrate strong scaling up to 1024 GPUs, analyze the shifting bottleneck from compute to communication at scale, and quantify the benefits of our pruned FFT optimization.

\subsection{Experimental Setup}

We evaluate our implementation on three supercomputers with distinct architectures: Alps at CSCS, LUMI at CSC Finland, and JUWELS Booster at JSC Germany.

\paragraph{Alps}
Alps is an HPE Cray EX supercomputer at the Swiss National Supercomputing Centre (CSCS). Every node contains four NVIDIA GH200 Grace Hopper Superchips, where each couples a 72-core NVIDIA Grace CPU (ARMv9 Neoverse V2) with an NVIDIA H100 GPU providing 96~GB HBM3 memory and 128~GB LPDDR5X CPU memory. The four GPUs within a node are fully connected via NVLink at 450~GB/s per direction, while inter-node communication uses the HPE Slingshot-11 interconnect at 25~GB/s per direction.

\paragraph{LUMI}
LUMI is an HPE Cray EX supercomputer at CSC in Kajaani, Finland. The GPU partition (LUMI-G) consists of 2978 nodes, each equipped with a single 64-core AMD EPYC 7A53 ``Trento'' CPU and four AMD MI250X GPUs. Each MI250X contains two Graphics Compute Dies (GCDs) with 64~GB HBM2e memory per die (128~GB per GPU, 512~GB per node). 
On LUMI, we count each MI250X GCD exposed by ROCm as one GPU device in the reported GPU counts. The two GCDs within each GPU communicate via Infinity Fabric at 200~GB/s per direction; inter-GPU links provide 50--100~GB/s per direction depending on topology. Each node connects to the HPE Slingshot-11 network via four NICs at 25~GB/s per direction each.

\paragraph{JUWELS Booster}
JUWELS Booster at J\"ulich Supercomputing Centre in Germany consists of 936 nodes. Each node contains two AMD EPYC 7402 ``Rome'' CPUs (24 cores each) and four NVIDIA A100 GPUs with 40~GB HBM2 memory each. The GPUs communicate within a node via NVLink3 at up to 300~GB/s per direction. Nodes connect through an HDR InfiniBand network at 25~GB/s per direction per link in a DragonFly+ topology, with four Mellanox ConnectX-6 adapters per node.

\paragraph{Software Environment}
Table~\ref{tab:software} summarizes the software configurations.
All NUFFT runs, both single-node and distributed, use Kokkos 5.0.2 and
HeFFTe 2.4.1.

\begin{table}[t]
\centering
\caption{Software environment on each system.}
\Description{A table listing the compiler, MPI implementation, and GPU SDK versions used on Alps, JUWELS Booster, and LUMI.}
\label{tab:software}
\begin{tabular}{@{}lccc@{}}
\toprule
 & Alps & JUWELS & LUMI \\
\midrule
Compiler & GCC 14.2 & GCC 13.3 & ROCm 6.3.4 \\
MPI & MPICH 8.1 & Open MPI 5.0 & MPICH 8.1 \\
GPU SDK & CUDA 12.9 & CUDA 12.6 & ROCm 6.3.4 \\
\bottomrule
\end{tabular}
\end{table}

\paragraph{Benchmarking Methodology}
All experiments are performed in double-precision arithmetic. FFT operations use cuFFT on
NVIDIA GPUs and rocFFT on AMD GPUs. Timings are measured on the host after
device synchronization. Unless otherwise noted, particles are initialized at
uniformly random positions. Each configuration uses five warm-up iterations
followed by 20 timed runs.

FFT operations use HeFFTe~\cite{krzhizhanovskaya_heffte_2020} configured with point-to-point reshaping, GPU-aware communication, and pencil decomposition. The default \texttt{MPI\_Alltoallv} reshaping exhibited suboptimal performance with Cray MPICH when GPU-aware MPI is enabled.

For comparisons with cuFINUFFT~\cite{shih_cufinufft_2021}, we use version 2.5.0, setting \texttt{gpu\_method=3} for spreading and otherwise default settings. We selected this configuration based on preliminary benchmarking experiments.

\subsection{Single-Node Performance} \label{subsec:single_gpu_results}
We first characterize single-GPU performance to establish baseline throughput and identify optimization opportunities before scaling to multiple nodes. All experiments in this subsection use $\rho = 10$ particles per cell and a $200^3$ grid unless stated otherwise. Throughput is reported as millions of nonuniform points processed per second (Mpts/s). We include comparisons against cuFINUFFT~\cite{shih_cufinufft_2021} on NVIDIA GPUs. cuFINUFFT targets NVIDIA hardware via CUDA and does not support AMD GPUs.

\subsubsection{Spreading and Interpolation Kernel Performance} \label{subsec:spread_interp_performance}

\begin{figure}[t]
    \centering
    \includegraphics[width=\linewidth]{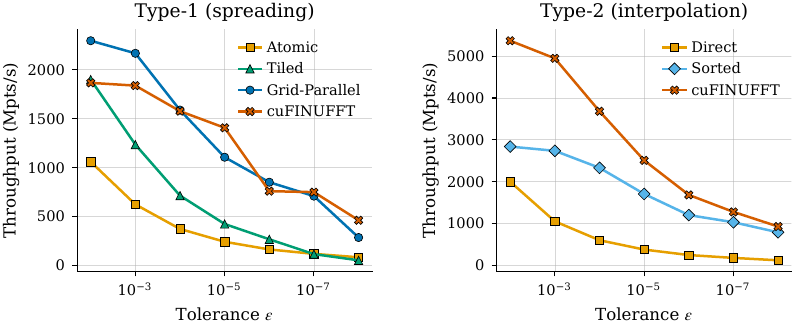}
    \caption{Kernel throughput as a function of tolerance on Alps (GH200)
    for a $200^3$ grid with $\rho = 10$. The comparison includes
    cuFINUFFT. Left: Type~1 spreading; right: Type~2 interpolation.
    Grid-Parallel spreading is competitive with cuFINUFFT at moderate
    tolerances and faster at loose tolerances, whereas cuFINUFFT achieves
    the highest interpolation throughput throughout.}
    \Description{Two line plots comparing kernel throughput on Alps as tolerance varies. The left plot shows Type 1 spreading, where Grid-Parallel performs best at loose tolerances and is close to cuFINUFFT at moderate tolerance. The right plot shows Type 2 interpolation, where cuFINUFFT is fastest across the tolerance range.}
    \label{fig:alps_kernels}
\end{figure}

\begin{figure}[t]
    \centering
    \includegraphics[width=\linewidth]{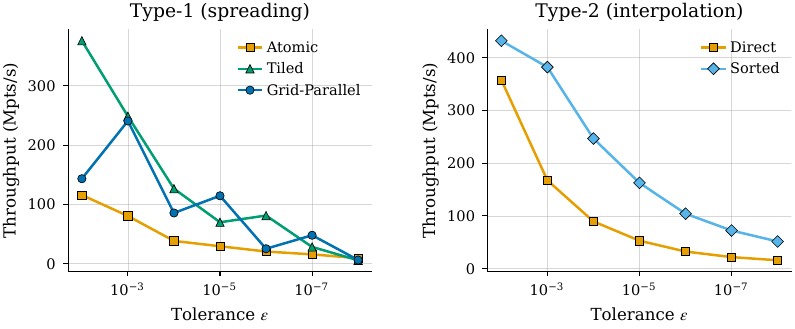}
    \caption{Kernel throughput versus tolerance on LUMI (MI250X), $200^3$ grid, $\rho = 10$. Both Tiled and Grid-Parallel have comparable performance overall. Sorted interpolation has superior throughput to Direct across all tolerances. cuFINUFFT does not support AMD GPUs.}
    \Description{Two line plots comparing kernel throughput on LUMI as tolerance varies. The left plot shows Type 1 spreading, where Tiled and Grid-Parallel are the strongest variants depending on tolerance. The right plot shows Type 2 interpolation, where Sorted interpolation is faster than Direct interpolation across the tolerance range.}
    \label{fig:lumi_kernels}
\end{figure}

The spreading and interpolation kernels dominate NUFFT runtime, so we begin by comparing kernel variants across tolerances. Figure~\ref{fig:alps_kernels} shows kernel throughput on Alps (GH200), including cuFINUFFT as a reference. Figure~\ref{fig:lumi_kernels} shows the corresponding results on LUMI (MI250X).

\paragraph{Spreading (Alps)}
Grid-Parallel achieves the highest spreading throughput on GH200 at looser tolerances, reaching 2297~Mpts/s at $\varepsilon = 10^{-2}$ and 1587~Mpts/s at $\varepsilon = 10^{-4}$. Compared to cuFINUFFT, Grid-Parallel is $1.2\times$ faster at $\varepsilon = 10^{-2}$ (2297 vs.\ 1866~Mpts/s) and reaches parity at $\varepsilon = 10^{-4}$ (1587 vs.\ 1577~Mpts/s). At tight tolerances ($\varepsilon \leq 10^{-7}$), cuFINUFFT shows better performance, reaching $1.6\times$ higher throughput at $\varepsilon = 10^{-8}$ (461 vs.\ 284~Mpts/s). Tiled spreading is consistently slower than both Grid-Parallel and cuFINUFFT. On NVIDIA H100-class GPUs, L1 cache and shared memory share unified physical storage~\cite{nvidia_hopper_tuning}, so explicit shared memory management in the Tiled kernel provides limited advantage over hardware-managed caching at high particle density. Atomic is consistently the slowest variant.

\paragraph{Interpolation (Alps)}
cuFINUFFT achieves higher interpolation throughput across all tolerances, ranging from $1.9\times$ at $\varepsilon = 10^{-2}$ (5371 vs.\ 2837~Mpts/s for Sorted) to $1.2\times$ at $\varepsilon = 10^{-8}$ (923 vs.\ 784~Mpts/s). Nsight Compute profiling attributes this gap primarily to integer-instruction overhead in the Kokkos-generated kernels relative to the hand-written CUDA code of cuFINUFFT. At tight tolerances, both implementations become memory-bound and the surplus integer work is hidden behind memory-access latency, narrowing the difference. Among our kernels, sorted interpolation dominates for $\varepsilon \leq 10^{-3}$: at $\varepsilon = 10^{-4}$, Sorted achieves 2328~Mpts/s versus 595~Mpts/s for Direct ($3.9\times$). Direct (unsorted) interpolation is competitive only at the loosest tolerance ($\varepsilon = 10^{-2}$), where the sort overhead exceeds the locality benefit.

\paragraph{Spreading (LUMI)}
On LUMI (Figure~\ref{fig:lumi_kernels}), the kernel ranking is less clear than on GH200. Tiled and Grid-Parallel deliver comparable throughput overall. Tiled is fastest at loose tolerances (376~Mpts/s at $\varepsilon = 10^{-2}$, $2.6\times$ Grid-Parallel) and at $\varepsilon = 10^{-6}$ (81 vs.\ 25~Mpts/s), while Grid-Parallel takes over at mid-to-tight tolerances such as $\varepsilon = 10^{-5}$ (114 vs.\ 70~Mpts/s) and $\varepsilon = 10^{-7}$ (48 vs.\ 28~Mpts/s). Atomic spreading is consistently the slowest variant, confirming that avoiding atomics is essential regardless of vendor.

\paragraph{Interpolation (LUMI)}
Sorted interpolation is optimal on LUMI for all tolerances, but absolute throughput is substantially lower than on GH200: at $\varepsilon = 10^{-4}$, Sorted achieves 247~Mpts/s on MI250X versus 2328~Mpts/s on GH200. Profiling reveals that the Sorted kernel allocates 100 registers with local memory spills on HIP, compared to only 32 registers on CUDA for identical source code, suggesting suboptimal code generation for index-heavy operations in the HIP compiler.

\paragraph{JUWELS Booster (A100)}
On JUWELS Booster, kernel rankings match Alps: Grid-Parallel leads for spreading (626~Mpts/s at $\varepsilon = 10^{-4}$) and Sorted for interpolation (481~Mpts/s). Absolute throughput is lower than on GH200, reflecting the generational difference between A100 and GH200. 

\subsubsection{Density Dependence}

\begin{figure}[t]
    \centering
    \includegraphics[width=\linewidth]{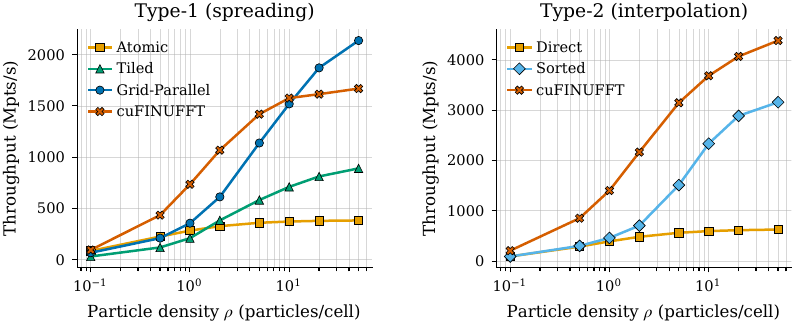}
    \caption{Kernel throughput versus particle density on Alps (GH200), $200^3$ grid, $\varepsilon = 10^{-4}$, including cuFINUFFT. Left: Type~1 (spreading). Right: Type~2 (interpolation). Grid-Parallel spreading surpasses cuFINUFFT at $\rho \geq 10$; cuFINUFFT interpolation remains faster across all densities.}
    \Description{Two line plots showing kernel throughput on Alps as particle density increases. The left plot shows Type 1 spreading, where Grid-Parallel improves with density and overtakes cuFINUFFT around density 10. The right plot shows Type 2 interpolation, where cuFINUFFT remains faster than Sorted and Direct interpolation across all densities.}
    \label{fig:density_sweep}
\end{figure}

Particle density significantly affects relative kernel performance. Figure~\ref{fig:density_sweep} shows throughput as a function of $\rho$ at fixed $\varepsilon = 10^{-4}$ on Alps (GH200).

For spreading, Grid-Parallel throughput scales nearly linearly with $\rho$, surpassing cuFINUFFT at $\rho \approx 10$ and reaching $1.3\times$ higher throughput at $\rho = 50$ (2138 vs.\ 1669~Mpts/s). This is because Grid-Parallel parallelizes over grid points and serializes particles within each tile: at higher density, each thread team processes more particles without additional atomic contention. cuFINUFFT's throughput saturates above $\rho = 10$. Tiled spreading scales similarly but has a lower throughput. Atomic plateaus at approximately 380~Mpts/s regardless of density, confirming that global atomic contention limits its throughput. At low density ($\rho < 10$), cuFINUFFT leads all methods, likely due to its optimized handling of sparse particle distributions.

For interpolation, cuFINUFFT maintains $1.4\times$ higher throughput than Sorted across all densities. Both methods scale with $\rho$ because higher density improves cache line utilization after sorting. Direct interpolation saturates at roughly 600~Mpts/s, confirming that the sorting step is essential for exploiting data locality at any substantial particle density.

These results show that the Grid-Parallel spreading kernel is well suited
to production densities, $\rho \geq 10$, in Particle-in-Fourier methods:
it matches or exceeds cuFINUFFT while remaining portable to AMD GPUs.

The LUMI results on MI250X GPUs show the same density-scaling trend. On
AMD, Tiled spreading is the fastest variant, reaching 125 and
456~Mpts/s at $\rho = 10$ and $\rho = 50$, respectively. Sorted
interpolation reaches 248 and 396~Mpts/s at the same densities. At fixed
density, absolute throughput is $9$--$13\times$ lower than on GH200,
consistent with the gaps identified in
Section~\ref{subsec:spread_interp_performance}.

\subsubsection{NUFFT Timing Breakdown}

\begin{figure}[t]
    \centering
    \includegraphics[width=\linewidth]{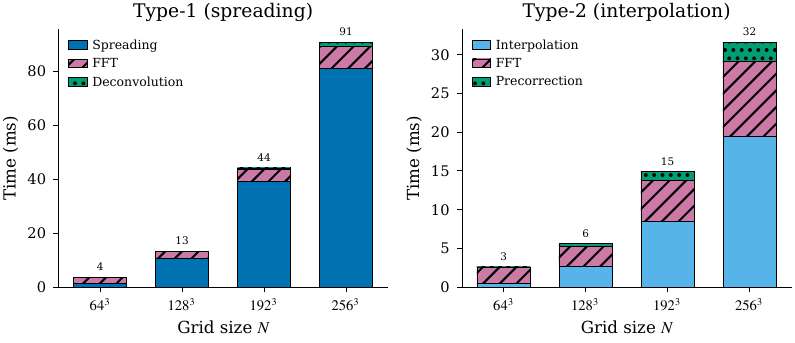}
    \caption{NUFFT time breakdown by component on Alps (GH200) for Type~1 (left) and Type~2 (right), $\rho = 10$, $\varepsilon = 10^{-4}$.}
    \Description{Two stacked bar charts showing the NUFFT runtime breakdown on Alps. The left chart shows Type 1, where spreading dominates for larger grids. The right chart shows Type 2, where interpolation is the largest component and the FFT is the second largest component.}
    \label{fig:nufft_breakdown}
\end{figure}

Figure~\ref{fig:nufft_breakdown} shows the time breakdown for Type~1 and Type~2 NUFFTs across grid sizes on Alps (GH200). Spreading dominates Type~1 runtime, accounting for 89\% at $N = 256$ (81~ms out of 91~ms). The FFT contributes only 9\% despite its $\mathcal{O}(N^3 \log N)$ complexity. This is because the FFT dispatches through HeFFTe to vendor-optimized libraries (cuFFT on NVIDIA, rocFFT on AMD) with highly regular memory access patterns, while the spreading kernel must handle irregular writes and atomic operations. Deconvolution is negligible at less than 2\%.

For Type~2, interpolation accounts for 62\% and the FFT for 31\% at $N = 256$. Interpolation is inherently faster than spreading (no atomic contention), giving the FFT a proportionally larger share. Pre-correction accounts for 8\%.

On LUMI (MI250X), the kernels consume an even larger fraction: spreading reaches 95\% of Type~1 runtime and interpolation 89\% of Type~2 at $N = 256$, because the slower kernel execution on AMD (see 
Section~\ref{subsec:spread_interp_performance}) amplifies the kernel-to-FFT ratio.

These results justify our focus on kernel optimization: improving spreading and interpolation performance directly impacts the dominant component of single-node NUFFT runtime.

\subsection{Pruned FFT Performance} \label{sub:pruned_fft_perf}                                                                                                                                                                       
\begin{figure}[t]                                  
      \centering 
      \includegraphics[width=\linewidth]{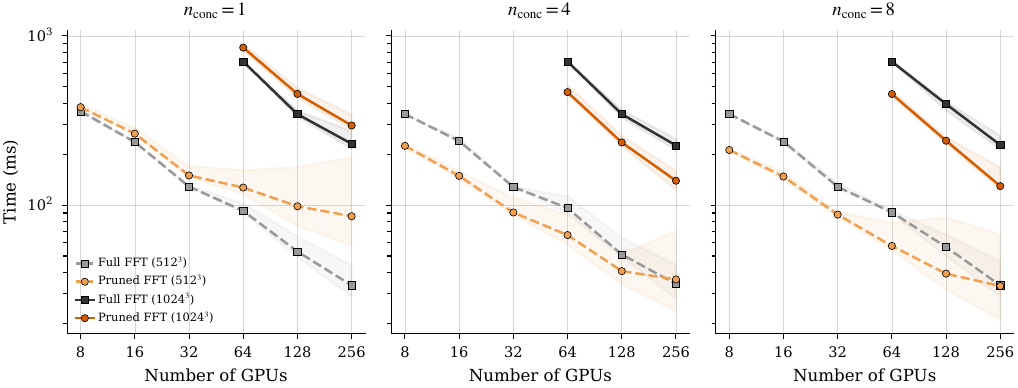}
      \caption{Pruned versus full FFT (forward) on Alps (GH200, Cray Slingshot) for output size $512^3$ and $1024^3$. Columns show different concurrency levels ($n_{\text{conc}}$).}
      \Description{Six line plots comparing pruned and full FFT execution time on Alps for two output grid sizes and three concurrency levels. For the larger grid, the pruned FFT is consistently faster than the full FFT. For the smaller grid, the pruned FFT shows higher variance at larger GPU counts. Increasing concurrency from 1 to 4 improves pruned FFT performance, while concurrency 8 gives diminishing returns.}
      \label{fig:pruned_fft}
  \end{figure}

  \begin{figure}[t]
      \centering
      \includegraphics[width=0.65\linewidth]{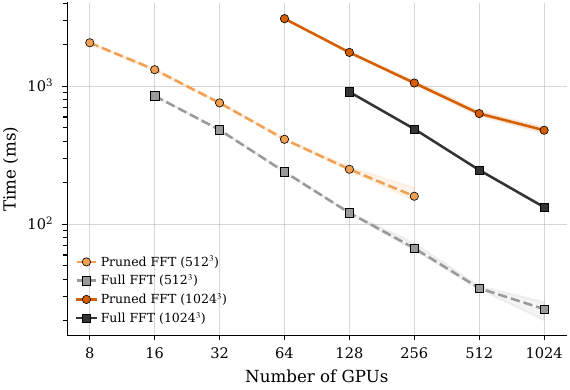}
      \caption{Pruned versus full FFT (forward) on JUWELS Booster (A100, InfiniBand) for output sizes $512^3$ and
  $1024^3$ with $n_{\mathrm{conc}} = 4$. The pruned FFT is consistently
   $2$--$2.5\times$ slower than the full FFT across all GPU counts.}
      \Description{Log-log line plot comparing pruned and full FFT execution time on JUWELS Booster for two output grid sizes. Across all GPU counts and both grid sizes, the full FFT is faster than the pruned FFT.}
      \label{fig:pruned_fft_juwels}
  \end{figure}

The pruned FFT (Section~\ref{sub:pruned_fft}) computes only the $N^3$ output modes required by the NUFFT rather than the full $(\sigma N)^3$ upsampled transform. This reduces both computation and communication by a constant. However, since the underlying distributed FFT is highly optimized, realizing this theoretical reduction in practice is not guaranteed---and, as we show below, whether the pruned FFT is beneficial depends critically on the MPI implementation and interconnect.

Figure~\ref{fig:pruned_fft} compares full and pruned FFTs on Alps
using GH200 GPUs, Slingshot-11, and Cray MPICH. The output grids are
$512^3$ and $1024^3$, corresponding to upsampled grids of $1024^3$ and
$2048^3$.

For the $1024^3$ grid, the pruned FFT achieves a $2\times$ speedup at
256 GPUs with low variance at $n_{\mathrm{conc}}=4$. The $512^3$ grid
shows higher variance at 128--256 GPUs, where outliers exceed the
median by up to $10\times$. Profiling shows increased MPI collective
time during sub-transforms. This is likely due to scheduling contention
when the runtime manages $\sigma^3$ concurrent communicators for
smaller local problem sizes.

Increasing concurrency from 1 to 4 sub-transforms improves performance by overlapping computation and communication. Beyond 4, benefits saturate and variance increases due to resource contention between concurrent streams.

On JUWELS Booster (A100, InfiniBand HDR, Open MPI), the pruned FFT performs significantly worse than the full FFT (Figure~\ref{fig:pruned_fft_juwels}). At 256~GPUs on the $512^3$ grid, the pruned FFT takes approximately 160~ms compared to 67~ms for the full FFT---a $2.4\times$ slowdown. This reversal persists across all GPU counts and both grid sizes. The pruned FFT decomposes the transform into $\sigma^3$ concurrent sub-transforms on independent sub-communicators. The efficiency of this approach depends heavily on how the MPI runtime handles concurrent communicators and collective operations. Open MPI over InfiniBand on JUWELS handles the $n_{\mathrm{conc}}$ concurrent sub-communicators substantially less efficiently than Cray MPICH over Slingshot-11 on Alps. LUMI (MI250X), which shares the same Slingshot-11 interconnect and Cray MPICH stack as Alps, exhibits similar pruned FFT behavior to Alps, confirming that the interconnect and MPI implementation---rather than the GPU architecture---are the determining factors.

This result highlights that algorithmic optimizations such as the pruned FFT are not universally portable: achieving optimal performance requires selecting the FFT strategy based on the target platform's communication infrastructure. We therefore expose the pruned FFT as an optional optimization. On systems with Cray MPICH and Slingshot, it provides benefits for larger grids. On systems with Open MPI and InfiniBand, the full FFT is preferable.

\subsection{Distributed Performance}

We evaluate strong scaling of the complete NUFFT on Alps (GH200), JUWELS Booster (A100), and LUMI (MI250X). The primary benchmark uses a $512^3$ grid with $\rho = 8$ particles per cell (1.07 billion points) and tolerance $\varepsilon = 10^{-4}$. We compare both FFT strategies introduced in Section~\ref{sub:pruned_fft}: the pruned FFT with concurrency $n_{\mathrm{conc}} = 4$ and the full FFT operating on the upsampled $(\sigma N)^3$ grid. We additionally report results on a $1024^3$ grid (8.6 billion points) where system resources permit.
\subsubsection{Grid Size $512^3$}

\begin{figure}[t]
    \centering
    \includegraphics[width=\linewidth]{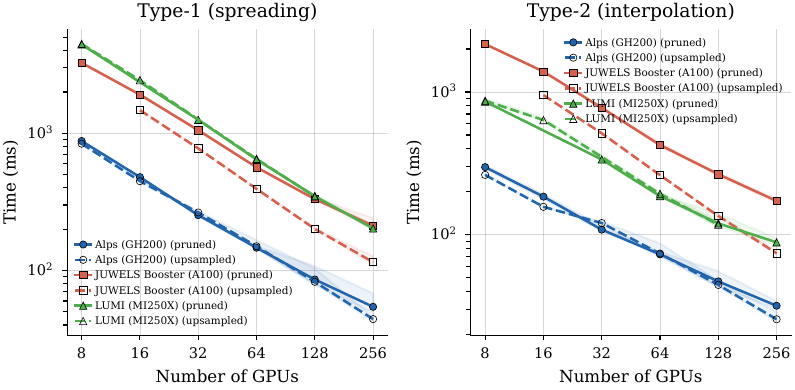}
    \caption{Strong scaling of the distributed NUFFT for $512^3$, $\rho=8$, $\varepsilon=10^{-4}$, on Alps, JUWELS Booster, and LUMI. Solid: pruned FFT ($n_{\mathrm{conc}}=4$); dashed: full FFT on the upsampled $1024^3$ grid. Shading shows min--max range.}
    \Description{Two log-log line plots showing strong scaling of the distributed NUFFT on Alps, JUWELS Booster, and LUMI for a 512 cubed grid. The left plot shows Type 1 and the right plot shows Type 2. Pruned and full FFT strategies are similar on Alps and LUMI, while the full FFT is consistently faster on JUWELS. Shaded bands indicate run-to-run variability.}
    \label{fig:nufft_scaling}
\end{figure}

\begin{figure}[t]
    \centering
    \includegraphics[width=\linewidth]{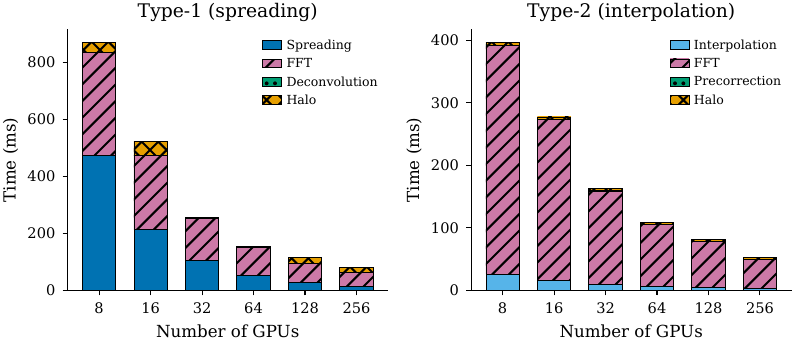}
    \caption{Component breakdown for the full-FFT NUFFT on Alps, $512^3$, $\rho=8$, $\varepsilon=10^{-4}$. Left: Type~1; right: Type~2.}
    \Description{Two stacked bar charts showing the NUFFT component breakdown on Alps for a 512 cubed grid using the full FFT. The left chart shows Type 1 and the right chart shows Type 2. As GPU count increases, spreading and interpolation decrease more rapidly than the FFT, making the FFT the dominant component at high GPU counts.}
    \label{fig:components}
\end{figure}

Figure~\ref{fig:nufft_scaling} compares the pruned and full FFT strategies across all three systems for the $512^3$ grid up to 256~GPUs. On Alps, both strategies achieve comparable end-to-end performance throughout the scaling range: the full FFT scales from 839~ms at 8~GPUs to 44~ms at 256~GPUs ($19\times$ speedup for $32\times$ resources, 60\% efficiency), while the pruned FFT reaches 54~ms at 256~GPUs (51\% efficiency). Notably, the full FFT is slightly faster despite the results in Section~\ref{sub:pruned_fft_perf}. We attribute this to the NUFFT's ability to overlap the FFT's communication with surrounding computation, reducing the practical benefit of the pruned FFT optimizations.

On JUWELS Booster, the full FFT is consistently and substantially faster, confirming the platform dependence established in Section~\ref{sub:pruned_fft_perf}. At 256~GPUs, Type~1 time drops from 212~ms (pruned) to 115~ms (full), a $1.8\times$ improvement. Type~2 shows a $2.3\times$ gap (172~ms versus 74~ms). The full FFT uses a single distributed transform on one communicator, avoiding the overhead of managing $n_{\mathrm{conc}}$ concurrent sub-communicators that penalizes the pruned FFT under Open MPI over InfiniBand. With the full FFT, JUWELS achieves 80\% parallel efficiency from 16 to 256~GPUs for Type~1.

On LUMI, both strategies perform comparably --- at 128~GPUs, the pruned and full FFT Type~1 times are within 1\% of each other (348~ms versus 347~ms). This matches our previous observations in 
Section~\ref{sub:pruned_fft_perf}. We note that HeFFTe on LUMI exhibited sporadic run failures due to unresolved errors in Cray MPICH's GPU-aware communication layer. These failures were independently observed and could not be resolved at publication time, limiting the number of valid data points on this system.

\paragraph{Component analysis}
Figure~\ref{fig:components} shows the component breakdown for the full NUFFT on Alps. The spreading kernel and FFT contribute comparably to Type~1 at 8~GPUs (54\% and 42\%, respectively). As the problem is distributed, spreading time decreases from 472~ms to 13~ms (8 to 256~GPUs, $35\times$ for $32\times$ resources), exhibiting super-linear scaling due to improved cache utilization at smaller local problem sizes. The FFT scales less favorably---from 363~ms to 49~ms ($7.4\times$)---because distributed FFTs require all-to-all communication bounded by interconnect bandwidth~\cite{czechowski_communication_2012,krzhizhanovskaya_heffte_2020}. This creates a shifting bottleneck: at 8~GPUs the FFT accounts for 42\% of Type~1. By 256~GPUs, it reaches 61\%. Halo exchange, which is negligible at low GPU counts (4\% at 8~GPUs), grows to 22\% at 256~GPUs as the surface-to-volume ratio of each subdomain increases. Type~2 is FFT-dominated across the entire range (92\% at 8~GPUs, 89\% at 256~GPUs), since interpolation is substantially less expensive than spreading.

The substantial contribution of the spreading kernel at moderate GPU counts validates the GPU kernel optimizations presented in Section~\ref{subsec:single_gpu_results}: spreading accounts for more than half of Type~1 runtime up to 16~GPUs, making kernel-level optimizations directly impactful on end-to-end performance. At higher GPU counts, the growing halo and FFT contributions indicate that communication optimizations become equally important.

\paragraph{Implications for higher-order methods}  
The shifting bottleneck has important implications for solver accuracy, although the effect is platform dependent. The spreading and interpolation kernels scale with the kernel width $w$, which increases for tighter tolerances and higher-order methods. The FFT cost is independent of $w$, 
and the halo-exchange cost grows only linearly with $w$, while the local particle--grid kernels grow as $w^d$. As the FFT and halo contributions grow at scale, the marginal cost of increasing $w$ decreases. On Alps at 256~GPUs, for instance, spreading accounts for only 17\% of the Type~1 runtime. This makes higher-accuracy configurations relatively inexpensive on systems where the FFT dominates. On LUMI, however, the particle--grid kernels remain a substantial part of the runtime, especially at tighter tolerances, so increasing $w$ still has a noticeable performance impact. Higher-order methods therefore become cheaper at scale only to the extent that communication-bound FFT costs dominate the overall runtime.

\subsubsection{Grid Size $1024^3$}

\begin{figure}[t]
    \centering
    \includegraphics[width=\linewidth]{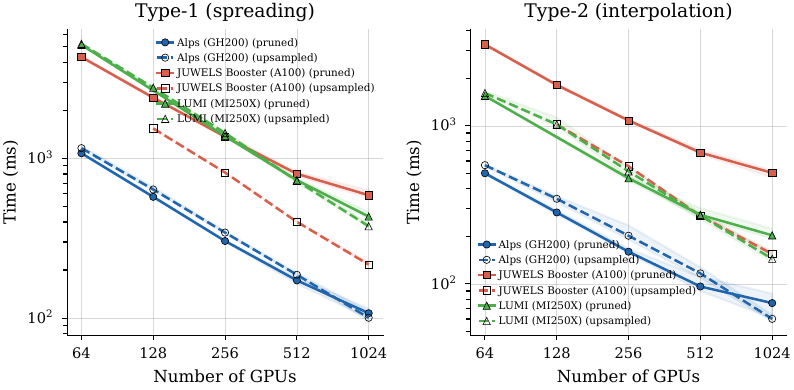}
    \caption{Strong scaling of the distributed NUFFT for a $1024^3$ grid, $\rho=8$, $\varepsilon=10^{-4}$. Solid: pruned FFT; dashed: full FFT. Shading shows min--max range.}
    \Description{Two log-log line plots showing strong scaling of the distributed NUFFT for a 1024 cubed grid. The left plot shows Type 1 and the right plot shows Type 2. On Alps, the pruned FFT is slightly faster than the full FFT. On JUWELS, the full FFT is clearly faster. On LUMI, the full FFT scales well up to 1024 GPUs.}
    \label{fig:nufft_scaling_1024}
\end{figure}

Figure~\ref{fig:nufft_scaling_1024} shows strong scaling for the $1024^3$ grid. On Alps, the pruned FFT scales from 1078~ms at 64~GPUs to 107~ms at 1024~GPUs ($10.1\times$ speedup for $16\times$ resources, 63\% efficiency). At this larger problem size, the pruned FFT is slightly faster than the full FFT on Alps---304~ms versus 343~ms at 256~GPUs---because the larger local problem sizes allow the computational savings of the pruned transform to outweigh the scheduling overhead observed at $512^3$.

On JUWELS, the full FFT again outperforms the pruned FFT, and the gap widens at higher GPU counts: $1.7\times$ at 256~GPUs, growing to $2.7\times$ at 1024~GPUs. The degradation worsens at scale because each GPU's smaller local volume amplifies the relative overhead of the concurrent sub-communicator FFTs, consistent with the HeFFTe-level root cause identified in Section~\ref{sub:pruned_fft_perf}. 

On LUMI, the full FFT scales from 5220~ms at 64~GPUs to 376~ms at 1024~GPUs ($13.9\times$ speedup for $16\times$ resources, 87\% efficiency)---the highest parallel efficiency among the three systems at this problem size. The pruned-FFT data on LUMI are limited to two GPU counts (64 and 256~GPUs) due to the Cray MPICH reliability issues described before. Where both strategies are available, they show comparable performance (${\approx}1380$~ms at 256~GPUs).

At this larger problem size, the component balance shifts less completely toward the FFT than at $512^3$: on Alps at 256~GPUs with the full FFT, spreading still accounts for 49\% of Type~1 (159~ms of 343~ms). The larger grid provides more computational work per GPU, keeping the kernel contribution higher across the scaling range. Nevertheless, the FFT share grows at the largest GPU counts, and production runs at this scale would still see diminishing marginal cost for tighter tolerances.

\section{Application: Particle-in-Fourier Schemes for Kinetic Plasma Simulations}

Particle-in-Fourier (PIF) schemes are spectral particle methods used in kinetic plasma simulations~\cite{mitchell_efficient_2019,shen_particle--fourier_2024,muralikrishnan_error_2025}. They offer high accuracy and favorable conservation and stability properties compared with the particle-in-cell schemes commonly used in these simulations. In a PIF simulation, the random particles (Monte Carlo samples) are first initialized with charges, positions, and velocities from the initial distribution. Each time step involves four operations:  
\begin{enumerate}
\item Scatter particle charges to Fourier space (Type~1 NUFFT). 
\item Solve the Poisson equation and compute the electric field in Fourier space.
\item Gather the electric field from Fourier space to the particle locations (Type~2 NUFFT).
\item Advance particle positions and velocities.
\end{enumerate}
Steps 1 and 3, the scatter and gather operations, dominate runtime. The main reason PIF schemes have not been pursued in production kinetic plasma simulations, despite their attractive numerical properties, is the lack of fast, distributed NUFFTs that can perform transforms involving billions of particles and Fourier modes.  
\begin{figure}[t]
    \centering
    \includegraphics[width=\columnwidth]{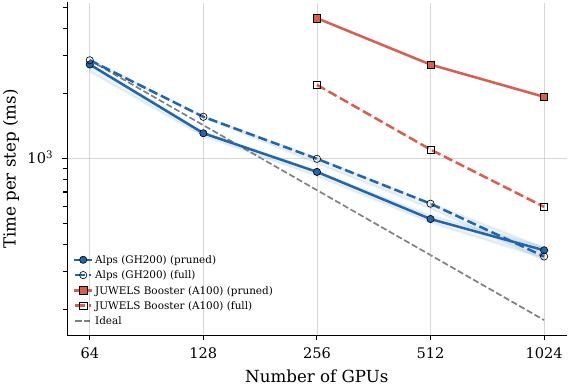}
    \caption{Strong scaling of PIF Landau damping at $1024^3$ modes, $\rho=8$, $\varepsilon=10^{-4}$. Solid: pruned FFT; dashed: full FFT. Shading shows min--max range.}
    \Description{Log-log line plot showing strong scaling of the Particle-in-Fourier Landau damping simulation at 1024 cubed modes on Alps and JUWELS Booster. Pruned and full FFT variants are close on Alps, while the full FFT is much faster on JUWELS. Shaded bands indicate run-to-run variability.}
    \label{fig:pif_scaling_1024}
\end{figure}

As an application of our distributed NUFFT, we perform 3D-3V PIF simulations of Landau damping, a classical benchmark problem in plasma physics~\cite{birdsall_plasma_1991,hockney_computer_1981}. We adopt initial conditions and parameters from~\cite{muralikrishnan2024scaling}. We consider the following initial distribution:
    \begin{align*}
         f(\xb, \vb, t=0) =& \frac{1}{\LRp{2\pi}^{3/2}} e^{-|\vb|^2/2} \LRp{1+\alpha\cos(k x)} \LRp{1+\alpha\cos(k y)} \\
                &\LRp{1+\alpha\cos(k z)},
     \end{align*}
in the domain $\LRs{0,L}^3$, where $L = 2\pi/k$ is the length in each dimension. We choose the parameters $k = 0.5$, $\alpha=0.05$, which correspond to weak Landau damping. Based on our initial distribution, the total electron charge is $Q_e = -L^3$. The NUFFT tolerance is chosen based on the accuracy and conservation requirements of the simulation and is coupled to the time-step size in explicit PIF schemes. We consider two tolerances: $\varepsilon=10^{-4}$ with time step $\Delta t=0.01$, and $\varepsilon=10^{-8}$ with time step $\Delta t=0.003125$. We use two problem sizes: $512^3$ Fourier modes ($1.07 \times 10^9$ particles) and $1024^3$ Fourier modes ($8.59 \times 10^9$ particles), both with $\rho=8$ particles per mode. Timing measurements use three warm-up steps followed by 10 timed steps to account for runtime initialization overhead. We use the fastest spreading method for the corresponding kernel width based on the kernel performance results in Section~\ref{subsec:single_gpu_results}. For Type~2, we use sorted interpolation. We evaluate both full and pruned FFTs to assess the pruned FFT performance in the full application context.

Figure~\ref{fig:pif_scaling_1024} shows strong scaling of the PIF simulation at $1024^3$ Fourier modes, $\rho=8$, with $\varepsilon=10^{-4}$, comparing pruned and full FFTs on Alps and JUWELS Booster. JUWELS Booster exhibits excellent scaling for full FFTs, achieving $91.9\%$ parallel efficiency from 256 to 1024~GPUs ($2197$~ms to $598$~ms). On Alps, full FFTs scale from $2856$~ms at 64~GPUs to $352$~ms at 1024~GPUs ($8.1\times$ speedup, $50.7\%$ parallel efficiency). Pruned FFTs on Alps are $13$--$16\%$ faster at 128--512~GPUs (e.g., $1311$~ms vs $1562$~ms at 128~GPUs), but the advantage diminishes at 1024~GPUs where both variants converge ($376$~ms vs $352$~ms), within the measurement variance. On JUWELS, however, pruned FFTs are $2$--$3.2\times$ slower than full FFTs at all GPU counts ($4475$~ms vs $2197$~ms at 256~GPUs), a consequence of HeFFTe performance degradation under the increased number of smaller distributed FFT calls that pruned transforms require. On LUMI, HeFFTe calls sporadically fail at scale for both full and pruned FFTs, and this instability is significantly amplified in the pruned case due to the higher number of FFT calls per time step, making reliable data collection infeasible. This issue was independently reproduced but not resolved before publication. As observed in Section~\ref{sub:pruned_fft_perf}, the pruned FFT optimization exhibits system-dependent characteristics that do not uniformly translate into application-level gains.

\begin{figure}[t]
    \centering
    \includegraphics[width=\columnwidth]{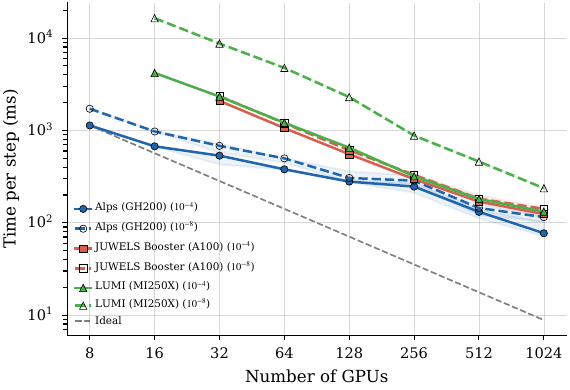}
    \caption{PIF strong scaling for $512^3$ modes, $\rho=8$, using full FFTs, at $\varepsilon=10^{-4}$ and $10^{-8}$.}
    \Description{Log-log line plot comparing Particle-in-Fourier time per step for two NUFFT tolerances on Alps, JUWELS Booster, and LUMI using a 512 cubed mode grid and the full FFT. JUWELS and LUMI scale well over most of the GPU range. The stricter tolerance adds little overhead on JUWELS but much larger overhead on LUMI. Alps shows weaker scaling at this problem size.}
    \label{fig:pif_tol_comparison}
\end{figure}

Figure~\ref{fig:pif_tol_comparison} compares the strong scaling performance of PIF at tolerances $\varepsilon = 10^{-4}$ and $\varepsilon = 10^{-8}$ using $512^3$ Fourier modes with $\rho = 8$, both employing full FFTs. JUWELS Booster and LUMI scale well at both tolerances. On JUWELS at $\varepsilon = 10^{-4}$, parallel efficiency from 32 to 512~GPUs is $76.9\%$ ($12.3\times$ speedup for $16\times$ resources), with pairwise efficiencies above $87\%$ at each doubling of GPUs. Beyond 512~GPUs, scaling degrades ($68.4\%$ pairwise efficiency from 512 to 1024~GPUs, $52.6\%$ overall from 32 to 1024~GPUs), as expected when the workload per GPU shrinks in strong scaling. On LUMI, efficiency from 16 to 512~GPUs is $72.4\%$, with pairwise efficiencies consistently above $88\%$, before similarly dropping to $68.5\%$ from 512 to 1024~GPUs. Scaling behavior at $\varepsilon=10^{-8}$ is similar on NVIDIA: on JUWELS, the stricter tolerance adds only $1.08$--$1.14\times$ overhead per step, since the distributed FFT---which dominates at higher GPU counts---is independent of the NUFFT tolerance for a fixed grid. However, on LUMI the cost of $\varepsilon=10^{-8}$ is $2.5$--$3.9\times$ higher than $\varepsilon=10^{-4}$, far exceeding the overhead on the NVIDIA platforms. Alps exhibits degraded scaling at $512^3$ ($14.7\times$ speedup from 8 to 1024~GPUs), which profiling suggests is associated with increased communication latency when
the NUFFT collectives interact with the application's particle communication. This effect is less pronounced at $1024^3$ where the larger per-GPU workload yields $50.7\%$ parallel efficiency (Figure~\ref{fig:pif_scaling_1024}).

\begin{figure}[t]
    \centering
    \includegraphics[width=\columnwidth]{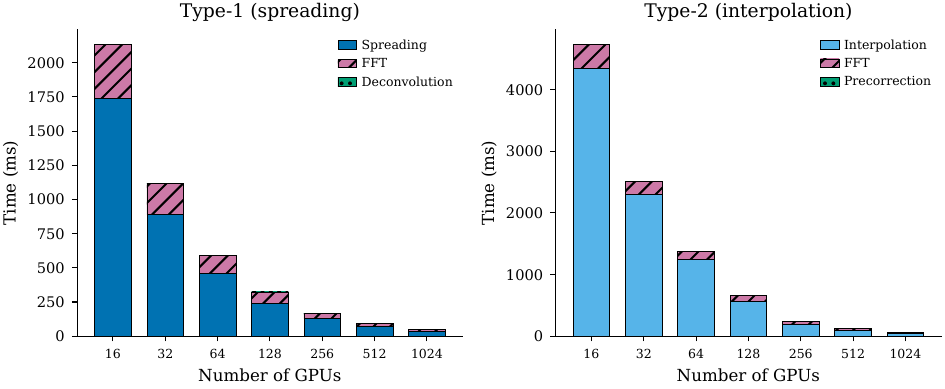}
    \caption{NUFFT component breakdown on LUMI at $512^3$, $\varepsilon=10^{-8}$, using full FFTs. Particle--grid kernels dominate even at high GPU counts.}
    \Description{Stacked bar chart showing the Type 1 and Type 2 NUFFT component breakdown on LUMI for a 512 cubed grid at the stricter tolerance. Spreading dominates Type 1 and interpolation dominates Type 2 across the GPU range, while the FFT is a smaller fraction of the runtime.}
    \label{fig:lumi_components_tol1e8}
\end{figure}

The disproportionate tolerance cost on LUMI is explained by the NUFFT component breakdown in Figure~\ref{fig:lumi_components_tol1e8}. At $\varepsilon=10^{-8}$, the spreading and interpolation GPU kernels dominate NUFFT time even at 1024~GPUs: spreading accounts for $65.5\%$ of Type~1 and interpolation for $66.7\%$ of Type~2 time. In contrast, at $\varepsilon=10^{-4}$ on the same platform, the distributed FFT dominates at high GPU counts ($69.7\%$ of Type~1, $59.1\%$ of Type~2 at 1024~GPUs). The spreading kernel alone is $4.0\times$ slower at $\varepsilon=10^{-8}$ compared to $\varepsilon=10^{-4}$ ($33.9$~ms vs $8.4$~ms at 1024~GPUs), and interpolation $3.0\times$ slower ($39.8$~ms vs $13.2$~ms). Since the FFT cost does not change with the NUFFT tolerance for a fixed grid, the additional kernel work shifts the bottleneck from communication to computation on LUMI. Profiling suggests two contributing factors: the wider interpolation kernel at $\varepsilon=10^{-8}$ increases per-point arithmetic, and the atomic operations central to the spreading kernel are significantly slower on the AMD MI250X. Combined with higher register spillage at wider kernel widths, as observed in Section~\ref{subsec:single_gpu_results}, this makes higher-accuracy transforms disproportionately expensive on AMD relative to NVIDIA.

\section{Conclusion}

We have presented, to the best of our knowledge, the first distributed
and performance-portable NUFFT for heterogeneous GPU systems.

Although the method is broadly applicable, it is motivated by
Particle-in-Fourier methods in computational physics~\cite{mitchell_efficient_2019,shen_particle--fourier_2024,muralikrishnan_error_2025,chen_fourier_2024}. In this setting, NUFFTs
must process billions of nonuniform points and Fourier modes across
thousands of GPUs at every time step.

We implemented and evaluated multiple spreading and interpolation strategies, identifying which kernel variants perform best under different accuracy requirements and particle densities. Grid-Parallel spreading reduces atomic contention and performs particularly well
at production particle densities, matching or exceeding the throughput of cuFINUFFT~\cite{shih_cufinufft_2021} on a single NVIDIA GPU, while additionally supporting AMD hardware. Sorted interpolation exploits spatial locality for similar gains but remains less performant than cuFINUFFT owing to integer-instruction overhead in the Kokkos-generated kernels.

We investigated a pruned FFT formulation for the distributed setting that computes only the required output modes rather than the full upsampled transform. While the pruned FFT reduces the theoretical computation and communication volume, our experiments show that it does not universally improve end-to-end performance: on systems with Cray MPICH and Slingshot (Alps, LUMI), the benefit is marginal and offset by reduced computation--communication overlap within the NUFFT pipeline, while on JUWELS Booster (Open MPI, InfiniBand) the pruned FFT is consistently slower than the full transform. This highlights that algorithmic optimizations targeting communication patterns are not performance-portable in practice and must be selected based on the target platform's MPI stack and interconnect.

We performed strong scaling experiments on three major European supercomputers: Alps (GH200), JUWELS Booster (A100), and LUMI (MI250X). At moderate tolerances, and especially on NVIDIA systems, the distributed FFT
 becomes the bottleneck at scale. In this regime, higher NUFFT accuracy
becomes relatively cheaper because the additional kernel work is amortized by
communication-bound FFT cost. On AMD MI250X, stricter tolerances remain
substantially more expensive because the particle--grid kernels continue to
dominate.

Finally, we demonstrate practical applicability through Particle-in-Fourier simulations of Landau damping with up to $1024^3$ Fourier modes and 8.6 billion particles. On JUWELS and LUMI at $512^3$, parallel efficiency remains above $72\%$ up to 512~GPUs, with pairwise efficiencies above $87\%$ at each doubling, before degrading at 1024~GPUs as expected in strong scaling. At $1024^3$, JUWELS achieves $91.9\%$ efficiency from 256 to 1024~GPUs. Comparing tolerances $\varepsilon=10^{-4}$ and $\varepsilon=10^{-8}$, the stricter accuracy adds only $1.1\times$ overhead on NVIDIA but up to $3.9\times$ on AMD, where the GPU kernels---rather than communication---remain the bottleneck. These results demonstrate that distributed NUFFTs enable kinetic plasma computations at resolutions previously inaccessible to spectral particle methods.

Our implementation is integrated into the IPPL framework and available as open source.

\paragraph{Limitations and Future Work}
Our HIP backend performance lags behind CUDA due to suboptimal register allocation in the compiler. Future work includes scaling studies to larger node counts and investigation of alternative distributed FFT algorithms to reduce the communication bottleneck at extreme scale.

\begin{acks}
We acknowledge access to Alps at the Swiss National Supercomputing Centre (CSCS), Switzerland, under the Paul Scherrer Institute's share with project ID psi07 and the PASC project with ID c41, which also provided access to PASC consultants. We acknowledge CSCS for awarding this project access to the LUMI supercomputer, owned by the EuroHPC Joint Undertaking and hosted by CSC (Finland) and the LUMI consortium. We also acknowledge the LUMI porting and optimization program and the AMD LUMI Hackathon (December 2025), hosted by CSCS, through which we gained performance insights and allocations on LUMI. Computing time on JUWELS Booster was provided under the projects CSTMA and Helmholtz Association's Initiative and Networking Fund on the HAICORE@FZJ partition. We thank John Biddiscombe, Radim Janalik, and Paul Bauer for valuable discussions.
\end{acks}

\section*{Availability}
IPPL is an open-source project released under the GNU General Public
License v3.0. The version of the framework used in this paper,
together with the raw benchmark data and the scripts that regenerate
all figures in this manuscript, is archived on Zenodo at
\url{https://doi.org/10.5281/zenodo.20057107}. The functionality
described here has been merged into the main IPPL repository, where
ongoing development continues, at
\url{https://github.com/IPPL-framework/ippl}.

\bibliographystyle{ACM-Reference-Format}
\bibliography{bib_final}

@article{dutt_fast_1993,
    title = {Fast {Fourier} {Transforms} for {Nonequispaced} {Data}},
    volume = {14},
    issn = {1064-8275, 1095-7197},
    url = {http://epubs.siam.org/doi/10.1137/0914081},
    doi = {10.1137/0914081},
    language = {en},
    number = {6},
    urldate = {2025-12-09},
    journal = {SIAM Journal on Scientific Computing},
    author = {Dutt, A. and Rokhlin, V.},
    month = nov,
    year = {1993},
    pages = {1368--1393},
}

@article{dutt_fast_1995,
    title = {Fast {Fourier} {Transforms} for {Nonequispaced} {Data}, {II}},
    volume = {2},
    issn = {10635203},
    url = {https://linkinghub.elsevier.com/retrieve/pii/S106352038571007X},
    doi = {10.1006/acha.1995.1007},
    language = {en},
    number = {1},
    urldate = {2026-04-29},
    journal = {Applied and Computational Harmonic Analysis},
    author = {Dutt, A. and Rokhlin, V.},
    month = jan,
    year = {1995},
    pages = {85--100},
}

@article{fessler_nonuniform_2003,
    title = {Nonuniform fast {Fourier} transforms using min-max interpolation},
    volume = {51},
    copyright = {https://ieeexplore.ieee.org/Xplorehelp/downloads/license-information/IEEE.html},
    issn = {1053-587X},
    url = {http://ieeexplore.ieee.org/document/1166689/},
    doi = {10.1109/TSP.2002.807005},
    language = {en},
    number = {2},
    urldate = {2026-04-29},
    journal = {IEEE Transactions on Signal Processing},
    author = {Fessler, J. A. and Sutton, B. P.},
    month = feb,
    year = {2003},
    pages = {560--574},
}

@article{arnold_comparison_2013,
    title = {Comparison of scalable fast methods for long-range interactions},
    volume = {88},
    copyright = {http://link.aps.org/licenses/aps-default-license},
    issn = {1539-3755, 1550-2376},
    url = {https://link.aps.org/doi/10.1103/PhysRevE.88.063308},
    doi = {10.1103/PhysRevE.88.063308},
    language = {en},
    number = {6},
    urldate = {2026-04-29},
    journal = {Physical Review E},
    author = {Arnold, Axel and Fahrenberger, Florian and Holm, Christian and Lenz, Olaf and Bolten, Matthias and Dachsel, Holger and Halver, Rene and Kabadshow, Ivo and G{\"a}hler, Franz and Heber, Frederik and Iseringhausen, Julian and Hofmann, Michael and Pippig, Michael and Potts, Daniel and Sutmann, Godehard},
    month = dec,
    year = {2013},
    pages = {063308},
}

@article{mitchell_efficient_2019,
    title = {Efficient {Fourier} basis particle simulation},
    volume = {396},
    issn = {00219991},
    url = {https://linkinghub.elsevier.com/retrieve/pii/S002199911930508X},
    doi = {10.1016/j.jcp.2019.07.023},
    language = {en},
    urldate = {2026-04-29},
    journal = {Journal of Computational Physics},
    author = {Mitchell, Matthew S. and Miecnikowski, Matthew T. and Beylkin, Gregory and Parker, Scott E.},
    month = nov,
    year = {2019},
    pages = {837--847},
}

@article{shen_particle--fourier_2024,
    title = {A particle-in-{Fourier} method with semi-discrete energy conservation for non-periodic boundary conditions},
    volume = {519},
    issn = {00219991},
    url = {https://linkinghub.elsevier.com/retrieve/pii/S0021999124006387},
    doi = {10.1016/j.jcp.2024.113390},
    language = {en},
    urldate = {2026-04-29},
    journal = {Journal of Computational Physics},
    author = {Shen, Changxiao Nigel and Cerfon, Antoine and Muralikrishnan, Sriramkrishnan},
    month = dec,
    year = {2024},
    pages = {113390},
}

@article{muralikrishnan_error_2025,
    title = {Error {Analysis} and {Parallel} {Scaling} {Study} of a {Parareal} {Parallel}-in-{Time} {Integration} {Algorithm} for {Particle}-in-{Fourier} {Schemes}},
    issn = {1064-8275, 1095-7197},
    url = {https://epubs.siam.org/doi/10.1137/24M1673097},
    doi = {10.1137/24M1673097},
    language = {en},
    urldate = {2026-03-20},
    journal = {SIAM Journal on Scientific Computing},
    author = {Muralikrishnan, Sriramkrishnan and Speck, Robert},
    month = oct,
    year = {2025},
    pages = {S311--S336},
}

@article{chen_fourier_2024,
    title = {A {Fourier} spectral immersed boundary method with exact translation invariance, improved boundary resolution, and a divergence-free velocity field},
    volume = {509},
    issn = {00219991},
    url = {https://linkinghub.elsevier.com/retrieve/pii/S0021999124002973},
    doi = {10.1016/j.jcp.2024.113048},
    language = {en},
    urldate = {2026-04-29},
    journal = {Journal of Computational Physics},
    author = {Chen, Zhe and Peskin, Charles S.},
    month = jul,
    year = {2024},
    pages = {113048},
}

@article{barnett_parallel_2019,
    title = {A {Parallel} {Nonuniform} {Fast} {Fourier} {Transform} {Library} {Based} on an ``{Exponential} of {Semicircle}'' {Kernel}},
    volume = {41},
    issn = {1064-8275, 1095-7197},
    url = {https://epubs.siam.org/doi/10.1137/18M120885X},
    doi = {10.1137/18M120885X},
    language = {en},
    number = {5},
    urldate = {2026-04-29},
    journal = {SIAM Journal on Scientific Computing},
    author = {Barnett, Alexander H. and Magland, Jeremy and af Klinteberg, Ludvig},
    month = jan,
    year = {2019},
    pages = {C479--C504},
}

@inproceedings{shih_cufinufft_2021,
    address = {Portland, OR, USA},
    title = {{cuFINUFFT}: a load-balanced {GPU} library for general-purpose nonuniform {FFTs}},
    copyright = {https://ieeexplore.ieee.org/Xplorehelp/downloads/license-information/IEEE.html},
    isbn = {978-1-6654-3577-2},
    shorttitle = {{cuFINUFFT}},
    url = {https://ieeexplore.ieee.org/document/9460591/},
    doi = {10.1109/IPDPSW52791.2021.00105},
    urldate = {2026-03-20},
    booktitle = {2021 {IEEE} {International} {Parallel} and {Distributed} {Processing} {Symposium} {Workshops} ({IPDPSW})},
    publisher = {IEEE},
    author = {Shih, Yu-hsuan and Wright, Garrett and Anden, Joakim and Blaschke, Johannes and Barnett, Alex H.},
    month = jun,
    year = {2021},
    pages = {688--697},
}

@article{potts_fast_2001,
    title = {Fast {Fourier} transforms for nonequispaced data: a tutorial},
    volume = {23},
    journal = {Mod. Sampl. theory},
    author = {Potts, Daniel and Steidl, Gabriele and Tasche, M.},
    month = jan,
    year = {2001},
    pages = {19--25},
}

@article{greengard_accelerating_2004,
    title = {Accelerating the {Nonuniform} {Fast} {Fourier} {Transform}},
    volume = {46},
    issn = {0036-1445, 1095-7200},
    url = {http://epubs.siam.org/doi/10.1137/S003614450343200X},
    doi = {10.1137/S003614450343200X},
    language = {en},
    number = {3},
    urldate = {2026-04-29},
    journal = {SIAM Review},
    author = {Greengard, Leslie and Lee, June-Yub},
    month = jan,
    year = {2004},
    pages = {443--454},
}

@article{keiner_using_2009,
    title = {Using {NFFT} 3---{A} {Software} {Library} for {Various} {Nonequispaced} {Fast} {Fourier} {Transforms}},
    volume = {36},
    issn = {0098-3500, 1557-7295},
    url = {https://dl.acm.org/doi/10.1145/1555386.1555388},
    doi = {10.1145/1555386.1555388},
    language = {en},
    number = {4},
    urldate = {2026-04-29},
    journal = {ACM Transactions on Mathematical Software},
    author = {Keiner, Jens and Kunis, Stefan and Potts, Daniel},
    month = aug,
    year = {2009},
    pages = {1--30},
}

@article{kunis_nonequispaced_2012,
    title = {The nonequispaced {FFT} on graphics processing units},
    volume = {12},
    copyright = {http://onlinelibrary.wiley.com/termsAndConditions\#vor},
    issn = {1617-7061, 1617-7061},
    url = {https://onlinelibrary.wiley.com/doi/10.1002/pamm.201210003},
    doi = {10.1002/pamm.201210003},
    language = {en},
    number = {1},
    urldate = {2026-04-29},
    journal = {PAMM},
    author = {Kunis, Susanne and Kunis, Stefan},
    month = dec,
    year = {2012},
    pages = {7--10},
}

@software{Polanco_NonuniformFFTs_jl_2024,
    author = {Polanco, Juan Ignacio},
    doi = {10.5281/zenodo.14637606},
    month = nov,
    title = {{NonuniformFFTs.jl}},
    url = {https://github.com/jipolanco/NonuniformFFTs.jl},
    version = {v0.6.7},
    year = {2024},
}

@article{lin_python_2018,
    title = {Python {Non}-{Uniform} {Fast} {Fourier} {Transform} ({PyNUFFT}): {An} {Accelerated} {Non}-{Cartesian} {MRI} {Package} on a {Heterogeneous} {Platform} ({CPU}/{GPU})},
    volume = {4},
    issn = {2313-433X},
    shorttitle = {Python {Non}-{Uniform} {Fast} {Fourier} {Transform} ({PyNUFFT})},
    url = {https://www.mdpi.com/2313-433X/4/3/51},
    doi = {10.3390/jimaging4030051},
    language = {en},
    number = {3},
    urldate = {2026-04-29},
    journal = {Journal of Imaging},
    author = {Lin, Jyh-Miin},
    month = mar,
    year = {2018},
    pages = {51},
}

@misc{muckley:20:tah,
    author = {Muckley, M. J. and Stern, R. and Murrell, T. and Knoll, F.},
    title = {{TorchKbNufft}: A High-Level, Hardware-Agnostic Non-Uniform Fast {Fourier} Transform},
    howpublished = {ISMRM Workshop on Data Sampling \& Image Reconstruction},
    year = {2020},
    note = {Source code available at https://github.com/mmuckley/torchkbnufft},
    url = {https://mmuckley.github.io/assets/publications/2020muckleytorchkbnufft.pdf},
}

@article{pippig_parallel_2013,
    title = {Parallel {Three}-{Dimensional} {Nonequispaced} {Fast} {Fourier} {Transforms} and {Their} {Application} to {Particle} {Simulation}},
    volume = {35},
    issn = {1064-8275, 1095-7197},
    url = {http://epubs.siam.org/doi/10.1137/120888478},
    doi = {10.1137/120888478},
    language = {en},
    number = {4},
    urldate = {2026-04-29},
    journal = {SIAM Journal on Scientific Computing},
    author = {Pippig, Michael and Potts, Daniel},
    month = jan,
    year = {2013},
    pages = {C411--C437},
}

@phdthesis{pippig_massively_2016,
    type = {{PhD} thesis},
    title = {Massively {Parallel}, {Fast} {Fourier} {Transforms} and {Particle}-{Mesh} {Methods}},
    author = {Pippig, Michael},
    school = {Technische Universit{\"a}t Chemnitz},
    address = {Chemnitz, Germany},
    year = {2016},
}

@inproceedings{muralikrishnan2024scaling,
    title = {Scaling and performance portability of the particle-in-cell scheme for plasma physics applications through mini-apps targeting exascale architectures},
    author = {Muralikrishnan, Sriramkrishnan and Frey, Matthias and Vinciguerra, Alessandro and Ligotino, Michael and Cerfon, Antoine J. and Stoyanov, Miroslav and Gayatri, Rahulkumar and Adelmann, Andreas},
    booktitle = {Proceedings of the 2024 SIAM Conference on Parallel Processing for Scientific Computing},
    pages = {26--38},
    year = {2024},
    doi = {10.1137/1.9781611977967.3},
    publisher = {Society for Industrial and Applied Mathematics},
    address = {Philadelphia, PA, USA},
    location = {Baltimore, MD, USA},
}

@misc{matthias_frey_ippl-frameworkippl_2024,
    title = {{IPPL}-framework/ippl: {IPPL}-3.2.0},
    copyright = {Creative Commons Attribution 4.0 International},
    shorttitle = {{IPPL}-framework/ippl},
    url = {https://zenodo.org/doi/10.5281/zenodo.10878166},
    doi = {10.5281/ZENODO.10878166},
    urldate = {2026-04-29},
    publisher = {Zenodo},
    author = {Frey, Matthias and Vinciguerra, Alessandro and Muralikrishnan, Sriramkrishnan and {Sonali} and {vmontanaro} and {Mohsen} and Adelmann, Andreas and {manuel5975p} and Schurk, Felix},
    month = mar,
    year = {2024},
}

@article{carter_edwards_kokkos_2014,
    title = {Kokkos: {Enabling} manycore performance portability through polymorphic memory access patterns},
    volume = {74},
    issn = {07437315},
    shorttitle = {Kokkos},
    url = {https://linkinghub.elsevier.com/retrieve/pii/S0743731514001257},
    doi = {10.1016/j.jpdc.2014.07.003},
    language = {en},
    number = {12},
    urldate = {2026-04-29},
    journal = {Journal of Parallel and Distributed Computing},
    author = {Carter Edwards, H. and Trott, Christian R. and Sunderland, Daniel},
    month = dec,
    year = {2014},
    pages = {3202--3216},
}

@incollection{krzhizhanovskaya_heffte_2020,
    address = {Cham},
    title = {{heFFTe}: {Highly} {Efficient} {FFT} for {Exascale}},
    volume = {12137},
    isbn = {978-3-030-50370-3 978-3-030-50371-0},
    shorttitle = {{heFFTe}},
    url = {https://link.springer.com/10.1007/978-3-030-50371-0_19},
    doi = {10.1007/978-3-030-50371-0_19},
    language = {en},
    urldate = {2026-04-29},
    booktitle = {Computational {Science} -- {ICCS} 2020},
    publisher = {Springer International Publishing},
    author = {Ayala, Alan and Tomov, Stanimire and Haidar, Azzam and Dongarra, Jack},
    editor = {Krzhizhanovskaya, Valeria V. and Z{\'a}vodszky, G{\'a}bor and Lees, Michael H. and Dongarra, Jack J. and Sloot, Peter M. A. and Brissos, S{\'e}rgio and Teixeira, Jo{\~a}o},
    year = {2020},
    note = {Series Title: Lecture Notes in Computer Science},
    pages = {262--275},
}

@article{barnett_aliasing_2021,
    title = {Aliasing error of the {$\exp(\beta\sqrt{1-z^2})$} kernel in the nonuniform fast {Fourier} transform},
    volume = {51},
    issn = {10635203},
    url = {https://linkinghub.elsevier.com/retrieve/pii/S1063520320300725},
    doi = {10.1016/j.acha.2020.10.002},
    language = {en},
    urldate = {2026-04-29},
    journal = {Applied and Computational Harmonic Analysis},
    author = {Barnett, Alex H.},
    month = mar,
    year = {2021},
    pages = {1--16},
}

@article{potts_uniform_2021,
    title = {Uniform error estimates for nonequispaced fast {Fourier} transforms},
    volume = {19},
    issn = {2730-5716, 2730-5724},
    url = {https://link.springer.com/10.1007/s43670-021-00017-z},
    doi = {10.1007/s43670-021-00017-z},
    language = {en},
    number = {2},
    urldate = {2025-12-09},
    journal = {Sampling Theory, Signal Processing, and Data Analysis},
    author = {Potts, Daniel and Tasche, Manfred},
    month = dec,
    year = {2021},
    pages = {17},
}

@article{potts_continuous_2021,
    title = {Continuous window functions for {NFFT}},
    volume = {47},
    issn = {1019-7168, 1572-9044},
    url = {https://link.springer.com/10.1007/s10444-021-09873-8},
    doi = {10.1007/s10444-021-09873-8},
    language = {en},
    number = {4},
    urldate = {2026-04-29},
    journal = {Advances in Computational Mathematics},
    author = {Potts, Daniel and Tasche, Manfred},
    month = aug,
    year = {2021},
    pages = {53},
}

@incollection{kaiser1966digitalfilters,
    author = {Kaiser, James F.},
    title = {Digital Filters},
    booktitle = {System Analysis by Digital Computer},
    editor = {Kuo, Franklin F. and Kaiser, James F.},
    publisher = {John Wiley and Sons},
    address = {New York},
    year = {1966},
    chapter = {7},
    pages = {218--285},
}

@book{osipov_prolate_2013,
    address = {Boston, MA},
    series = {Applied {Mathematical} {Sciences}},
    title = {Prolate {Spheroidal} {Wave} {Functions} of {Order} {Zero}: {Mathematical} {Tools} for {Bandlimited} {Approximation}},
    volume = {187},
    copyright = {https://www.springernature.com/gp/researchers/text-and-data-mining},
    isbn = {978-1-4614-8258-1 978-1-4614-8259-8},
    shorttitle = {Prolate {Spheroidal} {Wave} {Functions} of {Order} {Zero}},
    url = {https://link.springer.com/10.1007/978-1-4614-8259-8},
    doi = {10.1007/978-1-4614-8259-8},
    language = {en},
    urldate = {2025-12-09},
    publisher = {Springer US},
    author = {Osipov, Andrei and Rokhlin, Vladimir and Xiao, Hong},
    year = {2013},
}

@book{birdsall_plasma_1991,
    title = {Plasma {Physics} via {Computer} {Simulation}},
    author = {Birdsall, C. K. and Langdon, A. B.},
    year = {1991},
    publisher = {IOP Publishing},
    address = {Bristol, UK},
    series = {The Adam Hilger Series on Plasma Physics},
    url = {https://ui.adsabs.harvard.edu/abs/1991ppcs.book.....B},
    urldate = {2026-04-29},
    note = {ADS Bibcode: 1991ppcs.book.....B},
}

@book{hockney_computer_1981,
    address = {New York, NY},
    title = {Computer {Simulation} using particles},
    isbn = {978-0-07-029108-9},
    language = {eng},
    publisher = {McGraw-Hill},
    author = {Hockney, Roger W. and Eastwood, J. W.},
    year = {1981},
}

@inproceedings{knoll_gpunufft_2014,
    title = {{gpuNUFFT} -- An Open Source {GPU} Library for {3D} Regridding with Direct {Matlab} Interface},
    author = {Knoll, Florian and Schwarzl, Andreas and Diwoky, Clemens and Sodickson, Daniel K.},
    booktitle = {Proceedings of the International Society for Magnetic Resonance in Medicine},
    volume = {22},
    pages = {4297},
    year = {2014},
    publisher = {International Society for Magnetic Resonance in Medicine},
    address = {Berkeley, CA, USA},
    location = {Milan, Italy},
    url = {https://archive.ismrm.org/2014/4297.html},
}

@article{yang_cuda-based_2013,
    title = {A {CUDA}-based reverse gridding algorithm for {MR} reconstruction},
    volume = {31},
    issn = {0730-725X},
    url = {https://www.sciencedirect.com/science/article/pii/S0730725X12002482},
    doi = {10.1016/j.mri.2012.06.038},
    number = {2},
    urldate = {2026-03-16},
    journal = {Magnetic Resonance Imaging},
    author = {Yang, Jingzhu and Feng, Chaolu and Zhao, Dazhe},
    month = feb,
    year = {2013},
    pages = {313--323},
}

@article{feng_cuda_2015,
    title = {{CUDA} accelerated uniform re-sampling for non-{Cartesian} {MR} reconstruction},
    volume = {26 Suppl 1},
    issn = {1878-3619},
    doi = {10.3233/BME-151393},
    language = {eng},
    journal = {Bio-Medical Materials and Engineering},
    author = {Feng, Chaolu and Zhao, Dazhe},
    year = {2015},
    pages = {S983--989},
}

@article{gai_more_2013,
    title = {More {IMPATIENT}: {A} {Gridding}-{Accelerated} {Toeplitz}-based {Strategy} for {Non}-{Cartesian} {High}-{Resolution} {3D} {MRI} on {GPUs}},
    volume = {73},
    issn = {0743-7315},
    shorttitle = {More {IMPATIENT}},
    url = {https://pmc.ncbi.nlm.nih.gov/articles/PMC3652469/},
    doi = {10.1016/j.jpdc.2013.01.001},
    number = {5},
    urldate = {2026-03-16},
    journal = {Journal of Parallel and Distributed Computing},
    author = {Gai, Jiading and Obeid, Nady and Holtrop, Joseph L. and Wu, Xiao-Long and Lam, Fan and Fu, Maojing and Haldar, Justin P. and Hwu, Wen-mei W. and Liang, Zhi-Pei and Sutton, Bradley P.},
    month = may,
    year = {2013},
    pages = {686--697},
}

@techreport{morton1966computer,
    author = {Morton, Guy M.},
    title = {A Computer Oriented Geodetic Data Base and a New Technique in File Sequencing},
    institution = {IBM Ltd.},
    address = {Ottawa, Ontario, Canada},
    year = {1966},
}

@article{markel_fft_1971,
    title = {{FFT} pruning},
    volume = {19},
    copyright = {https://ieeexplore.ieee.org/Xplorehelp/downloads/license-information/IEEE.html},
    issn = {0018-9278},
    url = {http://ieeexplore.ieee.org/document/1162205/},
    doi = {10.1109/TAU.1971.1162205},
    language = {en},
    number = {4},
    urldate = {2026-03-12},
    journal = {IEEE Transactions on Audio and Electroacoustics},
    author = {Markel, J.},
    month = dec,
    year = {1971},
    pages = {305--311},
}

@article{sorensen_efficient_1993,
    title = {Efficient computation of the {DFT} with only a subset of input or output points},
    volume = {41},
    copyright = {https://ieeexplore.ieee.org/Xplorehelp/downloads/license-information/IEEE.html},
    issn = {1053587X},
    url = {http://ieeexplore.ieee.org/document/205723/},
    doi = {10.1109/78.205723},
    number = {3},
    urldate = {2026-03-12},
    journal = {IEEE Transactions on Signal Processing},
    author = {Sorensen, H. V. and Burrus, C. S.},
    month = mar,
    year = {1993},
    pages = {1184--1200},
}

@manual{nvidia_hopper_tuning,
  author       = {{NVIDIA Corporation}},
  title        = {{NVIDIA Hopper Tuning Guide}},
  subtitle     = {Tuning CUDA Applications for Hopper GPU Architecture},
  year         = {2026},
  organization = {NVIDIA Corporation},
  version      = {13.2},
  url          = {https://docs.nvidia.com/cuda/hopper-tuning-guide/index.html},
  urldate      = {2026-04-29},
  note         = {Last updated April 2, 2026},
}

@inproceedings{czechowski_communication_2012,
    address = {San Servolo Island, Venice, Italy},
    title = {On the communication complexity of {3D} {FFTs} and its implications for {Exascale}},
    isbn = {978-1-4503-1316-2},
    url = {https://dl.acm.org/doi/10.1145/2304576.2304604},
    doi = {10.1145/2304576.2304604},
    language = {en},
    urldate = {2026-03-20},
    booktitle = {Proceedings of the 26th {ACM} International Conference on Supercomputing},
    publisher = {ACM},
    author = {Czechowski, Kenneth and Battaglino, Casey and McClanahan, Chris and Iyer, Kartik and Yeung, P.-K. and Vuduc, Richard},
    month = jun,
    year = {2012},
    pages = {205--214},
}

\end{document}